# Multiphysics Continuous Shape Optimization of the TAP Reactor Components

Muhammad Ramzy Altahhan[1], Lynn Munday[2], Yousry Azmy[1]

[1]Department of Nuclear Engineering, North Carolina State University, Raleigh, NC 27695
[2]Idaho National Laboratory, Idaho Falls, ID 83352
mraltahh@ncsu.edu

## INTRODUCTION

The Transatomic Power (TAP) reactor has an unusual design for a molten salt reactor technology, building upon the foundation laid by the Molten Salt Reactor Experiment (MSRE). This design introduces three key modifications to enhance efficiency and compactness [1]: a revised fuel salt composition, an alternative moderator material, and moderator pins surrounded by the molten salt fuel. By replacing the MSRE's LiF-BeF2-ZrF4-UF4 mixture with LiF-UF4, the TAP reactor increases uranium concentration from 0.9 mol% to 27 mol% while maintaining a low melting point due to removal of the BeF2 and ZrF4 carrier salts. Additionally, replacing the graphite moderator with zirconium hydride (ZrH) results in a more compact core design, yielding a higher power density. These adaptations enable the TAP reactor to operate using commercially available low-enriched uranium (LEU) up to 5% enrichment, marking an advancement in the design of molten salt reactors – that traditionally use highly-enriched uranium [1]. Unlike traditional solid-fueled reactors that rely on excess positive reactivity at the beginning of life, the TAP concept employs a dynamic approach. The core's design, featuring a cylindrical geometry with square assemblies of moderator rods surrounded by flowing fuel salt, provides flexibility in adjusting the moderator-to-fuel ratio during operation – using movable moderator rods – further adding criticality control capability in addition to the control rods system.

Shape optimization of the core can play a crucial role in enhancing performance and efficiency. By applying multiphysics continuous shape optimization techniques to key components, such as the unit cells of the TAP reactor or its moderator assemblies, we can fine-tune the reactor's geometry to achieve optimal performance in key physics like neutronics and thermal hydraulics. We explore this aspect using the optimization module in the Multiphysics Object Oriented Simulation Environment (MOOSE) framework which allows for multiphysics continuous shape optimization [2]. The results reported here illustrate the benefits of applying continuous shape optimization in the design of nuclear reactor components and can help in extending the TAP reactor's performance.

## THEORY

The continuous shape optimization process aims to find the optimal geometric shape of a component based on performance criteria quantified by an objective function. This approach is popular in various industries – including nuclear systems – for improving efficiency and performance. The process begins from our discretization of our computational domain into nodes and elements, which are grouped into blocks and bounded by sidesets. The optimization algorithm modifies the position of nodes on selected sidesets to minimize an objective function. The problem is formulated as: Find the set of displacements $D$ that minimizes $O(\phi, D, \mathcal{N}_d)$. Here, $O$ is the objective function, $\phi$ represents the physics variables, and $\mathcal{N}_d$ is the set of displaced nodes resulting from applying the displacement set $D$ to the nondisplaced nodes $\mathcal{N}$.

To address the challenges of high dimensionality of the search space and potential element inversion, the displacement of nodes is constrained by a set of trial functions:

$$\vec{d}_j = \sum_{m=1}^{M} a_m \eta_m(\vec{n}_j), \qquad (1)$$

where $a_m$ are expansion coefficients and $\eta_m$ are trial functions with $\vec{n}_j$ being the coordinate vector of the $j$-th node. By utilizing Eq. (1), we have replaced the optimization problem over all $\mathcal{N}$ to an optimization problem over all $a_m$, $m = 1, .., M \ll \mathcal{N}$. To prevent element inversion, all nodes are allowed to move in response to the movement of optimized nodes, governed by:

$$\nabla \cdot \nabla \vec{d} = 0$$
$$\vec{d}(r) = \sum_{m=1}^{M} a_m \vec{\eta}_m(\vec{r}) \text{ on } \mathcal{S}$$
$$\vec{d}(r) = 0 \text{ on } \partial\Omega_e, \qquad (2)$$

where $\partial\Omega_e$ is the set of sidesets with nodes not allowed to displace (e.g., a potentially fixed outer boundary of the domain), $\mathcal{S}$ is the set of sidesets with nodes allowed to displace, and $\vec{r} = (x, y)$. This mesh displacement is known as Laplacian smoothing. The shape of a unit cell is shown in Figure 1 alongside the mesh used. The trial functions $\vec{\eta}$ we used in this analysis is judged by the circular shape of the moderator cell and are given by:

$$\vec{\eta} = \begin{pmatrix} \cos(\tan^{-1} y/x) \\ \sin(\tan^{-1} y/x) \end{pmatrix}, \qquad (3)$$

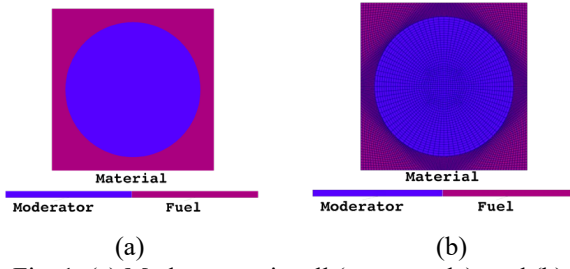

(a)        (b)

Fig. 1. (a) Moderator unit cell (not to scale), and (b) Moderator unit cell mesh used in the optimization (5,444 elements and 5,569 nodes).

which makes the displacements then be given as:

$$\vec{d} = \delta R_{mod}\vec{\eta} \quad \text{on } \mathcal{S} \in \partial\Omega, \qquad (4)$$

here $\delta R_{mod}$ is an optimization parameter representing the change in the moderator rod radius, and $\mathcal{S} \cup \partial\Omega_e = \partial\Omega$, the full boundary. This problem is solved on a dedicated MOOSE subapp which will provide the newly displaced mesh (as a solution of the problem in Eq. (2)) that will be used in the optimization process. The specification of sidesets that are allowed to deform (i.e., $\mathcal{S}$) and any fixed sidesets (i.e., $\partial\Omega_e$) is done in this subapp. This approach, implemented using MOOSE's displaced mesh capabilities, allows for optimization of complex geometries while maintaining mesh quality.

## MODEL DESCRIPTION

The TAP reactor has two basic unit cells, each a square 3 cm to the side: (1) a single moderator ZrH rod surrounded by fuel salt as depicted in Figure 1; and (2) a 3 cm cell of pure fuel salt. The ZrH rod has a radius of 1.15 cm, surrounded by a SiC cladding 0.10 cm thick. This unit cell configuration creates an arrangement where the space between the moderator rods is filled with liquid fuel salt. The difference between the pitch 3 cm and the total moderator rod diameter – 2.5 cm – leaves a gap of approximately 0.50 cm for the fuel salt to flow around each moderator rod in the first unit cell design which is used for the shape optimization. The TAP reactor core has 268 assemblies (67 per quadrant), with around 333 moderator unit cells per quadrant and with each assembly consisting of a 5 × 5 array of these unit cells. With a thermal power output of 1250 MWth, and a core radius of 1.55 m and an active fuel height of 300 cm, the reactor generates an average of 55.1 $W_{th}/cm^3$. This amounts to 496 $W_{th}/cm$ considering the unit cell area of 9 $cm^2$. This design allows for significant energy output from a relatively small core potentially improving reactor economics and siting flexibility, however, this high-power density also presents challenges in thermal management and fuel flow dynamics. The moderator unit cell was modeled in Serpent [3] considering a 5% enriched fuel salt with isotopic composition given in Table I.

Table I. Material Composition of the Serpent Model [1]

| Fuel Salt (5.01 g/cm³) | | Moderator (5.66 g/cm³) | |
|---|---|---|---|
| Isotope | Weight fr. | Isotope | Weight fr. |
| U-235 | 3.1100E-02 | Zr-90 | 4.9793E-01 |
| U-238 | 5.9090E-01 | Zr-91 | 1.0980E-01 |
| Li-7 | 4.8358E-02 | Zr-92 | 1.6967E-01 |
| Li-6 | 2.4180E-06 | Zr-94 | 1.7569E-01 |
| F-19 | 3.2964E-01 | Zr-96 | 2.8908E-02 |
| | | H-1 | 1.8007E-02 |
| | | H-2 | 4.1389E-06 |

## METHODOLOGY

We used the MOOSE framework optimization module to optimize the moderator unit cell of the TAP reactor shown in Figure 1a. The objective function $O$ used – built around the postprocessor system in MOOSE – is given by:

$$O(k_{eff}, T_m)$$
$$= \begin{cases} w \cdot T_m - k_{eff} \\ (w \cdot T_m - k_{eff}) \times 10, & \text{if } \begin{matrix} R_{mod} > 1.4\ cm \\ R_{mod} < 1.1\ cm \end{matrix} \end{cases} \quad (3)$$

By minimizing $O$, Eq. (3) maximizes the multiplication factor $k_{eff}$ while minimizing the maximum temperature in the unit cell, $T_m$, under a weighting function $w$, set to $10^{-3}$ to balance the contributions from $k_{eff}$ and $T_m \in [900 - 1000\ K]$. $T_m$ can also be used as a constraint considering the temperature-limited ZrH moderators due to the increase of both hydrogen mobility and thermal stresses driven cracks in the $700 - 800$ °C range ($1000 - 1100$ K). However, in the current study and assumptions, we do not reach this critical temperature hence we do not apply such a constraint on $T_m$. The value of $k_{eff}$ and the domain's temperature, needed for computing the maximum temperature in the domain as well as to provide feedback to the neutronics module, is computed by direct simulation of the domain's physics. We solved the multigroup neutron diffusion equation using the diffusion solver of Griffin [5]; and we solved thermal hydraulics using the heat transfer module of MOOSE. We used Serpent to prepare the cross sections. We parametrized the cross sections considering both the fuel temperature and the moderator radius, while the moderator temperature was kept at 600 K. The library points are given in Table II, including the initial values used for T and $R_{mod}$ in the optimization (shown in boldface). ISOXML tool from Griffin was used to generate the XML cross sections library from Serpent's results files for each run. The Serpent simulation, for the initial values given in Table II, yielded a $k_{eff}$ equal to , while Griffin using the neutron diffusion solver yielded Comparison between the deterministic neutronics solver and

Serpent's stochastic Monte Carlo simulation was 1002 pcm using two groups, and 565 pcm using 4 energy groups – given in Table II – hence we chose the 4 groups to perform the optimization. We performed a mesh refinement study that verified the mesh used in Figure 2 is adequate for the optimization process, e.g., 0.1 pcm difference in $k_{eff}$ is observed for the next refinement level.

Table II. Cross Section Library Points and Group Structure. Initial Nominal Values are bolded.

| Library points | | Energy Group Structure | |
|---|---|---|---|
| $T_{fuel}$ (K) | $R_{mod}$ (cm) | Group | Upper Bound (MeV) |
| 300 | 1.10 | 1 | 2.0000E+01 |
| 600 | 1.15 | 2 | 2.2310E+00 |
| | 1.20 | | |
| **900** | **1.25** | 3 | 4.9790E-01 |
| | 1.30 | 4 | 2.4788E-02 |
| 1200 | 1.35 | | |
| | 1.40 | 5 | 5.5308E-03 |
| | | 6 | 7.4850E-04 |
| | | | 1.0000E-11 |

$$O(k_{eff}, T_m)$$
$$= \begin{cases} w \cdot T_m - k_{eff} \\ (w \cdot T_m - k_{eff}) \times 10, & \text{if } \begin{array}{l} R_{mod} > 1.4 \ cm \\ R_{mod} < 1.1 \ cm \end{array} \end{cases} \quad (3)$$

In Eq. (3) a factor of 10 penalty constraint is applied to $O$ when the moderator rod radius ($R_{mod}$) exceeds the nuclear data library limits, given in Table II, effectively eliminating the possibility from consideration in the search since it is not included in the library. The direct relationship between $R_{mod}$ and mesh displacement is evident, however, only the mesh displacements are managed through MOOSE's reporter system. We implemented a custom MOOSE object, *ReporterValuePostprocessor*, which allows displacement values to be used as postprocessor quantities when needed for objective function penalty. In addition to the constraints system and bounded optimization provided by the Toolkit for Advanced Optimization (TAO) [4] - available through PETSc and utilized by MOOSE for continuous shape optimization - users have multiple options to apply, either individually or in combination, ensuring a consistent and constrained optimization process.

We used the gradient-free Nelder-Mead optimization algorithm from TAO for this single parameter optimization problem. However, considering the scope of the study, this algorithm was adequate. The default parameters of the Nelder-Mead optimizer were used; only the algorithm convergence parameter was reduced to 1E-08 from 1E-05 for a robust and extended optimization process..

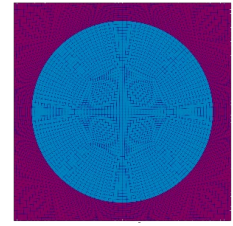

Fig. 2. Moderator unit cell mesh used in the optimization (7584 elements).

## RESULTS AND DISCUSSION

Starting by only considering the fuel temperature feedback on the cross sections in the moderated unit cell, the optimization process increases the moderator radius to ~1.32 cm. This entails an increased moderator to fuel ratio; the moderator rod almost completely fills the unit cell, and the increased moderation increases $k_{eff}$ by ~108.4 pcm. The maximum fuel temperature was also reduced by around ~10 K. The temperature distribution is shown in Figure 3; the thermal flux (Group 4) is shown in Figure 4 peaking at the rod's center and smallest at the unit cell's corners where the liquid fuel salt is most abundant.

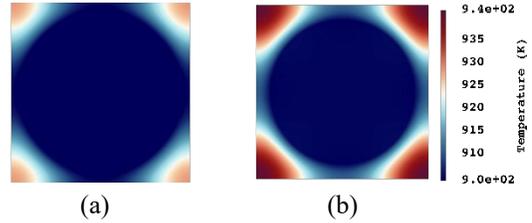

     (a)               (b)

Fig. 3. Moderator unit cell temperature distribution: (a) Optimized, and (b) Initial configuration

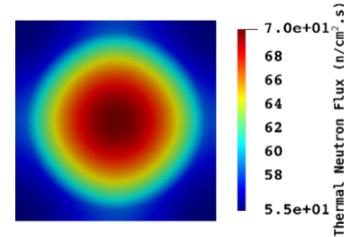

Fig. 4. Optimized moderator unit cell thermal neutrons flux (Group 4).

The temperature distribution is as expected considering that the heat is generated only in the fuel; the model's symmetry also plays a role in this distribution. The fuel's *subchannel* between four adjacent moderator rods is larger than the subchannel between any two adjacent rods resulting in more fuel salt in these regions. Furthermore, the heat transfer distance between two adjacent diagonally separated moderator rods is larger than the distance between two Cartesian-neighbor rods.

The heat conduction in the liquid fuel salt is negligible ($\sim 1$ W/mK) compared to the heat conduction of ZrH ($\sim 18$ W/mK) and shows a constant behavior with temperature [1]. It is worth noting that the detailed thermal conductivity behavior of LiF-UF4 salts is not available in the literature; no experimental value can be found for validation and the only value based on molecular dynamics simulations had a large uncertainty ($1.21 \pm 0.91$ W/mK) [5]. The convection heat transfer coefficient was computed considering a Nusselt number value of 3.61 (assuming laminar flow and square duct subchannel) and a hydraulic diameter equal to the 3 cm pin pitch, resulting in a value of $\sim 121$ W/m$^2$K.

The *ReporterValuePostprocessor* class we implemented allows us to get elements from the optimization process as quantitative values to use in parametrizing the cross sections. For instance, the moderator rod's radius can be used as a parameter to parametrize the cross sections when building the library in Serpent as seen in Table II. Griffin only allows field variables when parametrizing the cross sections; any scalar variable postprocessor value cannot be used for parameterization and will result in a runtime error. The solution was to project the scalar postprocessor into an auxiliary field variable value using MOOSE's *MultiAppPostprocessorInterpolationTransfer* object. This object interpolates postprocessor values from different subapps and projects the computed value into a field variable; ideal for our case since we only have one postprocessor which is the moderator radius and hence its correct value as computed by the postprocessors will be projected. We verified that the new auxiliary variable shows the correct value of the moderator radius postprocessor.

Repeating the optimization but this time accounting for the moderator radius feedback on the cross sections, showed a reduction in the moderator radius to $\sim 1.20$ cm. Negligible effect on $k_{eff}$ was observed showing a mere increase of $\sim 31$ pcm. The maximum temperature increased by approximately 6 K, despite being included in the objective function for minimization. These results could indicate either that the optimizer has reached a local minimum and would require the optimization process to be tested using a different optimizer from Nelder-Mead that allows rigorous approaches for escaping local minima, or that the current configuration is indeed the most optimal configuration given the simulation and the physics solved.

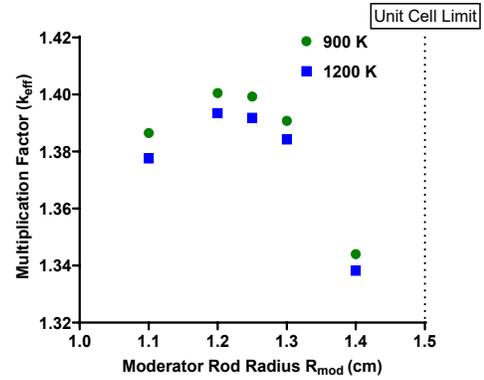

Fig. 6. $k_{eff}$ behavior against the moderator radius for 900 K and 1200 K cross sections using Serpent.

To test this hypothesis, we plotted the $k_{eff}$ computed using Serpent against the parametrized moderator radius for the 900 K and the 1200 K temperatures in Figure 6. Apparently, the moderator radius effect on $k_{eff}$ is not *monotonic* and it shows a parabolic behavior in which clearly a maximum point can be found around 1.2 cm, which the optimizer predicted, and amounts to the correct fuel to moderator ratio in the moderator unit cell that maximizes $k_{eff}$.

## CONCLUSIONS AND FUTURE WORK

This study demonstrated the application of MOOSE's continuous shape optimization module to enhance the TAP reactor's moderator unit cell design, focusing on maximizing $k_{eff}$ while minimizing the peak temperature. Key findings include an initial optimization that increased the moderator radius to $\sim 1.32$ cm, leading to higher keff (+108.4 pcm) and reduced maximum fuel temperature (-10 K) when considering only fuel temperature feedback. Incorporating moderator radius feedback on cross-sections resulted in a reduced optimal moderator radius ($\sim 1.2$ cm) with minimal impact on keff (+31 pcm) and a small increase in maximum temperature (+6 K). Analysis revealed a non-monotonic, parabolic dependence of keff on moderator radius, with an optimal point around 1.2 cm aligning with the optimizer's prediction. Future work will explore hierarchical optimization at the assembly level, considering the arrangement and configuration of multiple unit cells, and will exercise combinatorial optimization to determine the optimal number and placement of moderator rods per assembly.

## ACKNOWLEDGMENTS

This research used funding received from the DOE Office of Nuclear Energy's Nuclear Energy University Programs under grant DE-NE0009308 project number 22-26770.